\documentclass{article}

\usepackage{cite,url,graphicx,color}
\usepackage{anysize}
\marginsize{1.1in}{1.1in}{0.99in}{1in}

\newcommand{\hide}[1]{}

\title{\vspace{-12mm}
Limits on Fundamental Limits to Computation}
\date{\vspace{-5mm}}
\author{\vspace{-6mm} Igor L. Markov, the University of Michigan (currently at Google)}

\begin{document}
\maketitle
\vspace{-4mm}
\begin{abstract}
 An indispensable part of our lives, \emph{computing} has also become essential to industries and governments. Steady improvements in computer hardware have been supported by periodic doubling of transistor densities in integrated circuits over the last fifty years.
 Such \emph{Moore scaling} now requires increasingly heroic efforts, stimulating research in alternative hardware and stirring controversy. To help evaluate emerging technologies and enrich our understanding of integrated-circuit scaling, we review fundamental limits to computation: in manufacturing, energy, physical space, design and verification effort, and algorithms.
 To outline what is achievable \emph{in principle} and \emph{in practice}, we recall how some limits were circumvented, compare loose and tight limits. We also point out that engineering difficulties encountered by emerging technologies may indicate yet-unknown limits.
\hide{To interpret loose limits as indications for feasible improvements, we propose to study \emph{iterated limits} and illustrate this approach by examples.}
\end{abstract}

\vspace{-2mm}
\section{Introduction}
\label{sec:intro}
\vspace{-2mm}
   Emerging technologies for computing promise to outperform conventional integrated circuits in  computation bandwidth or speed, power consumption, manufacturing cost, or form factor \cite{Cavin12,Chien13}. However, razor-sharp focus on any one nascent technology and its benefits sometimes neglects serious limitations or discounts ongoing improvements in established approaches. To foster a richer context for evaluating emerging technologies, we review limiting factors and salient trends in computing that determine what is achievable \emph{in principle} and \emph{in practice}. Several fundamental limits remain substantially loose, possibly indicating viable opportunities for emerging technologies. To clarify this uncertainty, we study \emph{limits on fundamental limits}.

{\bf Universal and general-purpose computers.}
   Viewing {\em clocks and watches} as early computers, it is easy to see the importance of
   {\em long-running calculations that can be repeated with high accuracy by mass-produced devices}.
   The significance of {\em programmable} digital computers became clear at least 200 years ago, as illustrated by {\em Jacquard looms} in textile manufacturing. However, the existence of {\em universal computers} that can efficiently simulate (almost) all other computing devices --- analog or digital --- was only articulated in the 1930s by Church and Turing (Turing excluded quantum physics when considering universality) \cite{Herken13}.\hide{Turing doubted that quantum physics can be efficiently simulated with conventional physics, but only in the 1980s did Manin \cite{Manin80} and Feynman \cite{Feynman86} suggest quantum computers.} Efficiency was studied from a theoretical perspective at first, but strong demand in military applications in the 1940s lead Turing and von Neumann to develop detailed hardware architectures for universal computers --- Turing's design (Pilot ACE) was more efficient, but von Neumann's was easier to program. The {\em stored-program architecture} made universal computers practical in the sense that a single computer design could be effective in many diverse applications. Such practical universality thrives $(i)$ in economies of scale in computer hardware and $(ii)$ among extensive software stacks. Not surprisingly, the most sophisticated and commercially successful computer designs and components, such as Intel and IBM CPUs, were based on the von Neumann's paradigm.
   The numerous uses and large markets of general-purpose chips, as well as the exact reproducibility of their results, justify the enormous capital investment in the design, verification and manufacturing of leading-edge integrated circuits. Today general-purpose CPUs power cloud server-farms and displace specialized (but still universal) mainframe processors in many supercomputers. Emerging universal computers based on field-programmable gate-arrays (FPGAs) and general-purpose graphics processing units (GPGPUs) outperform CPUs in some cases, but their efficiencies remain complementary to those of CPUs. The success of deterministic general-purpose computing manifests in the convergence of diverse functionalities in portable inexpensive smartphones. After steady improvement, general-purpose computing displaced entire industries (newspapers, photography, etc) and launched new applications (video conferencing, GPS navigation, online shopping, networked entertainment, etc) \cite{Andreesen11}.
   Application-specific integrated circuits (ASICs) streamline input-output and networking, or optimize functionalities previously performed by general-purpose hardware. They speed up biomolecular simulation 100-fold \cite{Padua11,Shaw13} and improve the efficiency of video decoding 500-fold \cite{Hameed11}, but require design effort with keen understanding of specific computations, impose high costs and financial risks, need markets where general-purpose computers lag behind, and often cannot adapt to new algorithms. Recent techniques for {\em customizable domain-specific computing} \cite{Cong11} offer better tradeoffs, while many applications favor the combination of {\em general-purpose hardware and domain-specific software}, including specialized programming languages \cite{Mernik05,Olukotun14} such as Erlang used in Whatsapp.

{\bf Limits as aids to evaluating emerging technologies.}
   Without sufficient history, we cannot extrapolate \emph{scaling laws} for emerging technologies, yet expectations run high. For example, new proposals for analog processors appear frequently (as illustrated by adiabatic quantum computers), but fail to address concerns about analog computing, such as its limitations on scale, reliability, and long-running error-free computation. General-purpose computers meet these requirements with digital integrated circuits (IC) and now command the electronics market.
   In comparison, quantum computers --- both digital and analog --- hold promise {\em only in niche applications} and do not offer {\em faster general-purpose computing} as they are no faster for {\em sorting} and other specific tasks \cite{Aaronson04,Jain10,Nielsen11}. In exaggerating the engineering impact of quantum computers, popular press missed this nuance. But in scientific research, building quantum computers may help simulating quantum-chemical phenomena
   and reveal new fundamental limits. Sections \ref{sec:space} 
   and \ref{sec:conc} discuss limits on emerging technologies.

{\bf Technology extrapolation versus fundamental limits.}
The scaling of commercial computing hardware regularly runs into formidable obstacles \cite{Cavin12}, but near-term technological advances often circumvent them. The International Technology Roadmap for Semiconductors (ITRS) \cite{ITRS} keeps track of such obstacles and possible solutions with a focus on frequently-revised consensus estimates. For example, consensus estimates initially predicted 10 GHz CPUs for the 45 nm technology node \cite{Sylvester00}, versus the 3-4 GHz range seen in practice. In 2004, the unrelated Quantum Information Science and Technology Roadmap \cite{Hughes04} forecast 50 physical qubits by 2012. Such optimism arose by assuming technological solutions long before they were developed and validated, and by overlooking important limits. The authors of \cite{Meindl95,Davis01} classify limits to device and interconnect as {\em fundamental, material, device, circuit}, and {\em system limits}. These categories define the rows of Table \ref{tab:limits}, and the columns reflect sections of this paper where specific limits are examined for tightness.
\hide{Section \ref{sec:new} discusses the possibility of discovering new limits to computation,
and Section \ref{sec:conc} summarizes our review.}

\begin{table}
\hspace{-7mm}
\includegraphics[width=5.8in]{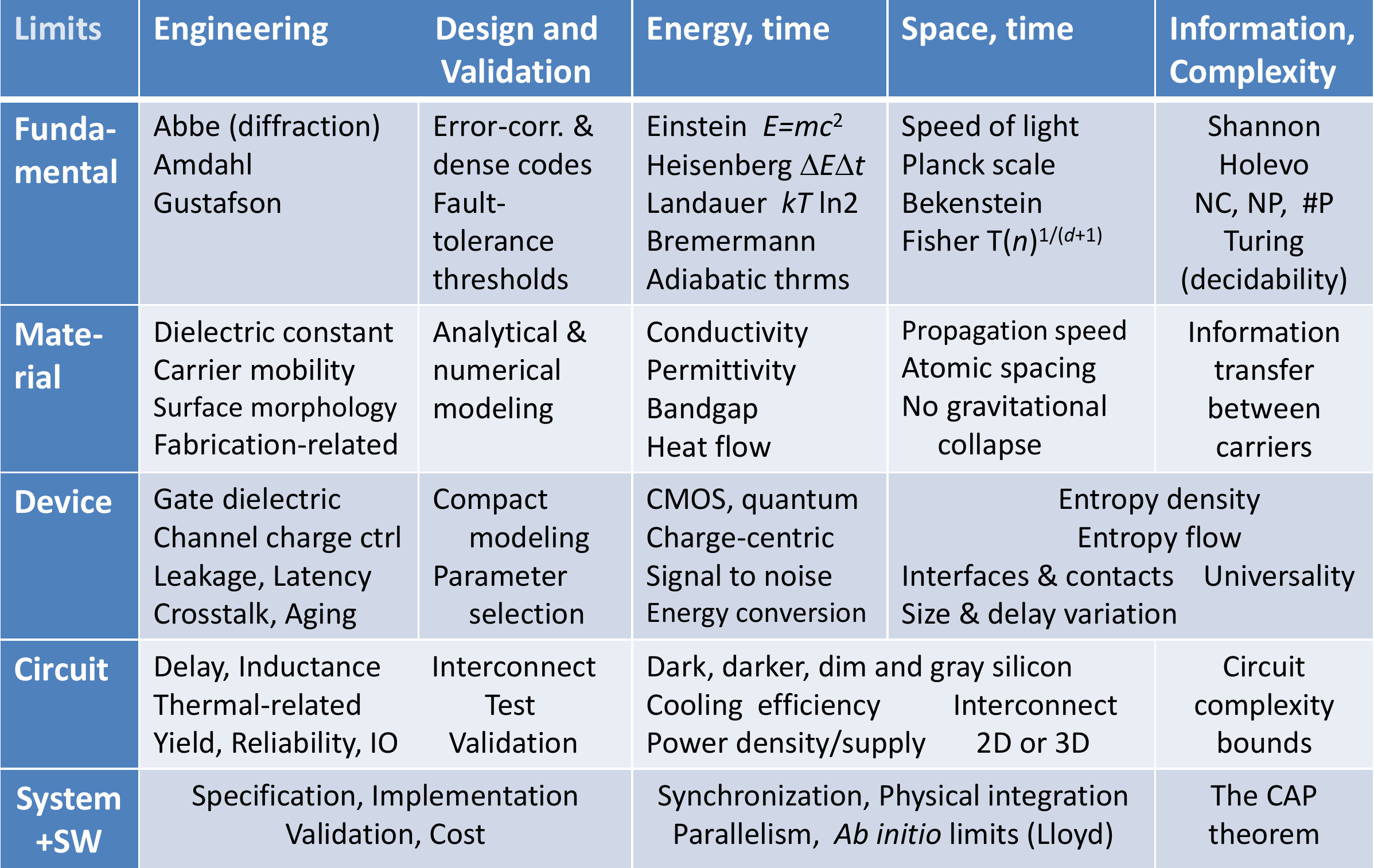}
\hspace{-3mm}
\hspace{-3mm}
\vspace{-3mm}
\parbox{13cm}{
\caption{
 \label{tab:limits} Some of the known limits to computation
 \cite{Aharonov61,Pendry83,Fisher88,Meindl95,Bohr95,Asenov98,Ho99,Blencowe00,Sylvester00,Lloyd00,Davis01,Wolf01,Hisamoto02,Vazirani02,
 ZhirnovCHB03,Saxena04,Kim05,Adya06,Bohr07,Borkar07,Fortnow09,Dreslinski10,Padua11,Nielsen11,
 CAP12,Shelar13,Taylor12,Esmaielzadeh13,Berut12,Sipser12,Markov13,ITRS,Naeemi14}.
}
}
\vspace{-1mm}
\end{table}

\vspace{-2mm}
\section{Engineering obstacles}
\label{sec:eng}
\vspace{-2mm}
  Engineering obstacles limit specific technologies and choices. For example, a key bottleneck today is IC manufacture, which packs billions of transistors and wires in several $cm^2$ of silicon with astronomically low defect rates. Layers of material are deposited on silicon and patterned with lasers, fabricating all circuit components simultaneously. Precision optics and photochemical processes ensure accuracy.

{\bf Limits on manufacturing.}
No account of limits to computing is complete without the Abbe diffraction limit: light with wavelength $\lambda$, traversing a medium with refractive index $\eta$, and converging to a spot with angle $\theta$ (perhaps, focused by a lens) creates a spot with diameter $d=\lambda/NA$, where
$NA=\eta \sin \theta$ is the numerical aperture. $NA$ reaches 1.4 for modern optics, so it would seem that semiconductor manufacturing is limited to feature sizes $\lambda/2.8$, hence ArF lasers with 193nm wavelength should not support photolithographic manufacturing of transistors with 65nm features. Yet, they supported {\em sub-wavelength lithography} for the 45nm-22nm technology nodes using {\em asymmetric illumination} and {\em computational lithography} \cite{MaA11}. Here one starts with optical masks that look like the intended image, but when the image gets blurry, alter masks by gently shifting edges to improve the image, possibly giving up the semblance between the two. Clearly, some limits are formulated to be broken!
Ten years ago, researchers demonstrated patterning of nanomaterials by live viruses \cite{Mazzola03}. Known virions exceed 20nm in diameter, whereas subwavelength lithography with 193nm-wavelength ArF laser recently extended to 14nm semiconductor manufacturing \cite{ITRS}. Hence, viruses and microorganisms are no longer at the forefront of semiconductor manufacturing. Extreme ultra-violet (X-ray) lasers have been energy-limited, but are improving. Their use requires changing refractive optics to reflective.
Additional progress in {\em multiple patterning} and {\em directed self-assembly} promises to support photolithography beyond the 10nm technology node. \hide{\cite{Merritt13}.}

\begin{figure}[tb]
\begin{center}
\vspace{-2mm}
\includegraphics[width=8.0cm]{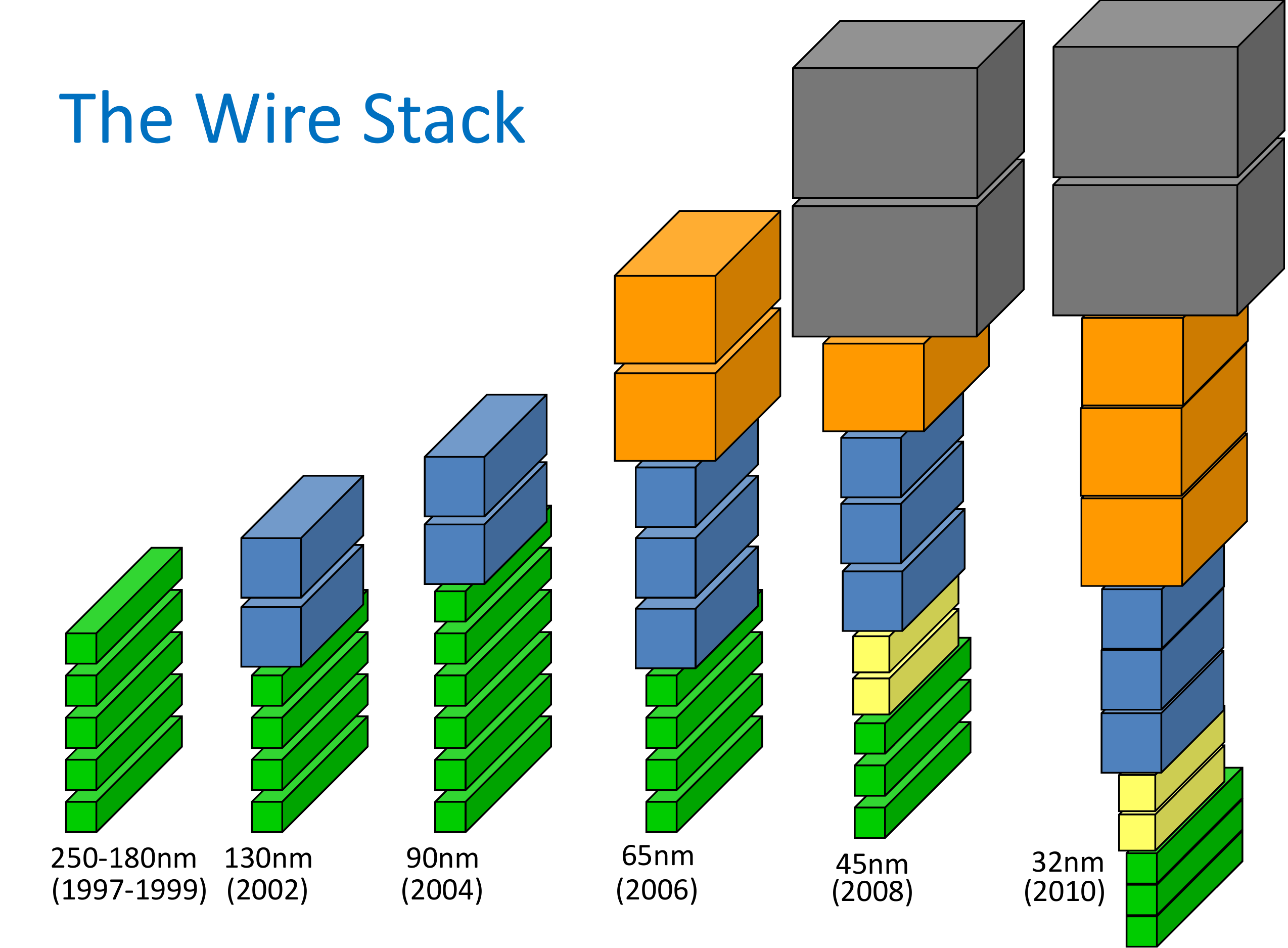}
\vspace{-2.5mm}
\caption{ \label{fig:wires} The evolution of metallic wire stacks from 1997 to 2010 by semiconductor technology nodes. {\sc Image credit: IBM Research} (modified).
}
\vspace{-2mm}
\end{center}
\end{figure}

{\bf Limits on individual interconnects.}
Despite the doubling of transistor density with Moore's law \cite{Moore65}, semiconductor integrated circuits (ICs) would not work without fast and dense interconnects. Metallic wires can be either fast or dense, but not both at the same time --- smaller cross-section increases electrical resistance, while greater height or width increase parasitic capacitance with neighboring wires (wire delay grows with $RC$). In 1995, an Intel researcher pointed out that {\em on-chip interconnect scaling} is the real limiter of high-performance ICs \cite{Bohr95}. The scaling of interconnect is also moderated by {\em electron scattering against rough edges of metallic wires} \cite{Davis01,Naeemi14}, inevitable with atomic-scale wires. Hence, IC interconnect stacks have evolved \cite{Sylvester00,Shelar13} from four equal-pitch layers in 2000 to 16 layers with pitches varying by 32 times, including a large amount of dense (thin) wiring and fast (thick) wires used for global on-chip communication (Figure \ref{fig:wires}). Aluminum and copper remain unrivaled for conventional interconnects and can be combined in short wires \cite{Naeemi14}; {\em carbon-nanotube} and {\em spintronic} interconnects are also evaluated in \cite{Naeemi14}. {\em Photonic waveguides} and {\em RF links} offer alternative IC interconnect \cite{Almeida04,Chang02}, but obey fundamental limits derived from Maxwell's equations, such as the {\em maximum propagation speed of EM waves} \cite{Davis01}. I/O links are limited by the perimeter or surface area of a chip, whereas chip capacity grows with area or volume, respectively.

\begin{figure}[t]
\begin{center}
\includegraphics[width=7.8cm]{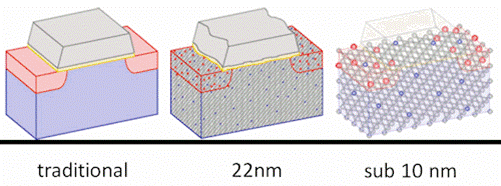}
\vspace{-3mm}
\caption{ \label{fig:xtor-atoms} As MOSFET transistors shrink, gate dielectric (yellow) thickness approaches several atoms (0.5nm at the 22nm technology node). Atomic spacing limits device density to 1 device/nm even for radical devices.
}
\end{center}
\vspace{-6mm}
\end{figure}

\begin{figure}[tb]
\begin{center}
\vspace{-1mm}
\includegraphics[width=7.8cm]{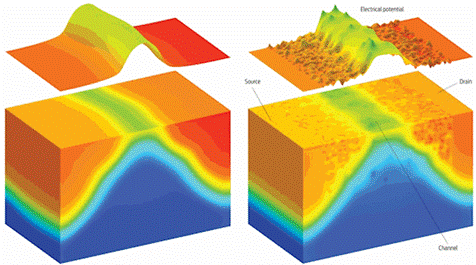}
\vspace{-3mm}
\caption{ \label{fig:xtor-field} As MOSFET transistors shrink, the shape of electric field departs from basic rectilinear models, and level curves become disconnected. Atomic-level manufacturing variations, especially for dopant atoms, start affecting device parameters, making each transistor slightly different \cite{Asenov98,Miranda12}. {\sc Image credit: Gold Standard Simulations}.
}
\end{center}
\vspace{-6mm}
\end{figure}

\begin{figure}[tb]
\vspace{-1mm}
\begin{center}
\includegraphics[height=3.7cm]{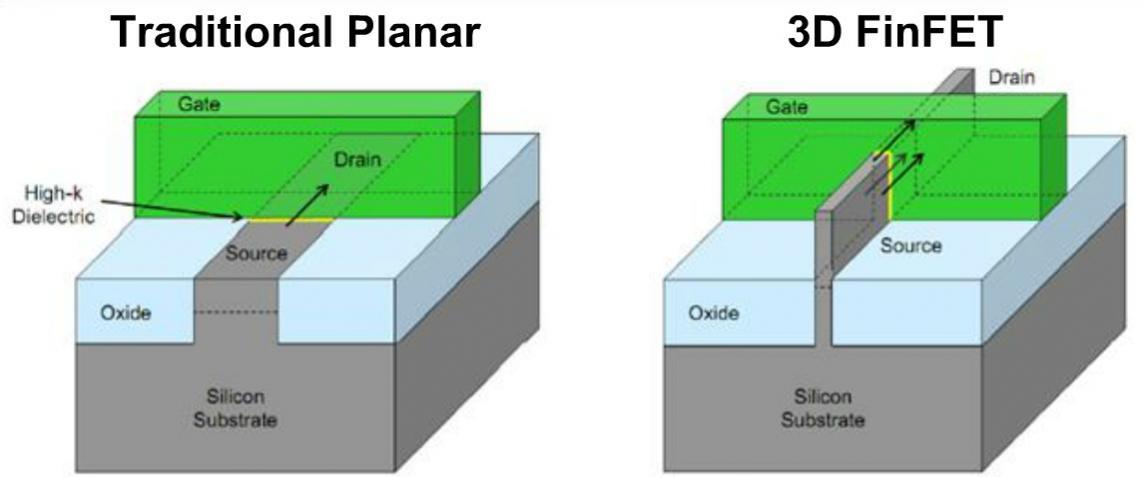}
\vspace{-3mm}
\caption{ \label{fig:xtor-fin} FinFET transistors possess much wider gate dielectric (surrounding the fin shape) than MOSFET transistors and can use multiple fins.
}
\vspace{-6mm}
\end{center}
\end{figure}

{\bf Limits on conventional transistors.}
Transistors are limited by their tiniest feature --- {\em the width of the gate dielectric}, --- which recently reached the size of several atoms (Figure \ref{fig:xtor-atoms}), creating problems: $(i)$ a few missing atoms could alter transistor performance, $(ii)$ manufacturing variation makes all transistors slightly different (Figure \ref{fig:xtor-field}), $(iii)$ electric current tends to leak through thin narrow dielectrics \cite{Meindl95}. \hide{Can transistors improve without thinner gate dielectric?} Instead of a {\em thinner} dielectric, transistors can be redesigned with {\em wider} dielectric layers \cite{Hisamoto02} that surround a {\em fin shape} (Figure \ref{fig:xtor-fin}). Such configurations improve the control of electric field, reduce current densities and leakage, and diminish process variations. Each transistor can use several fins, extending transistor scaling by several generations. Semiconductor manufacturers adopted FinFETs for upcoming technology nodes. 
One step further, in \emph{tunneling transistors} \cite{Seabaugh13} a gate wraps around the channel to control tunnelling rate.

{\bf Limits on design effort.}
\hide{Integrated circuit design has always pushed the boundaries of practical feasibility.}
In the 1980s, Mead and Conway formalized IC design using a regular grid, enabling automated layout through algorithms. But resulting optimization problems remain hard, and heuristics are only good enough for practical use. Besides frequent algorithmic improvements, each technology generation alters circuit physics and requires new CAD software. The cost of design has doubled in a few years, becoming prohibitive for ICs with limited market penetration \cite{ITRS}.
 Emerging technologies, such as FinFETs and high-K dielectrics, circumvent known obstacles using forms of design optimization. Therefore, reasonably tight limits should account for potential future optimizations. Low-level technology enhancements, no matter how powerful, are often viewed as one-off improvements, in contrast to architectural redesigns that affect many processor generations.
Between technology enhancements and architectural redesigns are global and local optimizations that alter ``the texture'' of IC design, such as {\em logic restructuring, gate sizing and device parameter selection}. Moore's law promises higher transistor densities, but some transistors are designed to be 32 times larger than others. Large gates consume greater power to drive long interconnects at acceptable speed and satisfy performance constraints.
Minimizing circuit area and power, subject to timing constraints (by configuring each logic gate to a certain size, threshold voltage, etc), is a hard but increasingly important optimization with a large parameter space. A recent convex optimization method \cite{Ozdal12} saved 30\% power in Intel chips, and the impact of such improvements grows with circuit size. Many aspects of IC design are being improved, continually raising the bar for technologies that compete with CMOS.

Completing new IC designs, optimizing and verifying them requires great effort and continuing innovation, e.g., the lack of scalable design automation is a limiting factor for analog ICs \cite{Rutenbar06,Rutenbar10}. In 1999, bottom-up analysis of digital IC technologies  \cite{Ho99,Sylvester00} outlined design scaling up to self-contained modules with 50K standard cells (each cell contains 1-3 logic gates), but further scaling was limited by global interconnect. In 2010, physical separation of modules became less critical, as large-scale placement optimizations assumed greater responsibility for IC layout and learned to blend nearby modules \cite{Markov12,Puri13}. In a general trend, powerful design automation \cite{Lavagno06} frees circuit engineers to focus on microarchitecture \cite{Puri13}, but increasingly relies on algorithmic optimization. Until recently, this strategy suffered significant losses in performance \cite{Chinnery04} and power \cite{Chinnery07} compared to ideal designs, but has now became both successful and indispensable due to rapidly increasing complexity of digital and mixed-signal electronic systems. Hardware and software must now be co-designed and co-verified, with software efforts increasing at a faster rate. {\em Platform-based design} combines high-level design abstractions with effective reuse of components and functionalities in engineered systems \cite{Vincentelli04}. Customizable domain-specific computing \cite{Cong11} and domain-specific programming languages \cite{Mernik05,Olukotun14} offload specialization to software running on reusable hardware platforms.

\section{Energy-time limits}
\label{sec:energy}
In predicting the main obstacles to improving modern electronics, the International Technology Roadmap for Semiconductors (ITRS) highlights {\em the management of system power and energy} as the dominant Grand Challenge \cite{ITRS}. The faster the computation, the more energy it consumes, but actual power-performance tradeoffs depend on the physical scale. While the ITRS, by its charter, focuses on near-term projections and IC design techniques, fundamental limits reflect available energy resources, properties of the physical space, power-dissipation constraints, and energy waste.

\subsection{Reversibility}
A 1961 result by Landauer \cite{Landauer61} shows that erasing one bit of information entails an energy loss $\geq kT \ln 2$ ({\em thermodynamic threshold}), where $k$ is the Boltzmann constant and $T$ is the temperature in Kelvin. This principle was validated empirically in 2012 \cite{Berut12} and seems to motivate {\em reversible computing} \cite{Bennett85}, where all input information is preserved, incurring additional costs. Formally speaking, zero-energy computation is prohibited by {\em the energy-time form of the Heisenberg uncertainty principle} ($\Delta t \Delta E \geq \hbar$): faster computation requires greater energy \cite{Aharonov61,Lloyd00}. However, recent work in applied superconductivity \cite{Ren11} demonstrates ``highly exotic'' {\em physically-reversible} circuits operating at 4$^{\circ}$K with energy dissipation below the thermodynamic threshold. They apparently fail to scale to large sizes, run into other limits, and remain no more practical than ``mainstream'' superconducting circuits and refrigerated  low-power CMOS circuits. Technologies that implement {\em quantum circuits} \cite{Monroe14} can {\em approximate} reversible Boolean computing, but currently do not scale to large sizes, are energy-inefficient at the system level, rely on fragile components, and require heavy fault-tolerance overhead \cite{Nielsen11}. Conventional ICs also do not help obtaining energy savings from reversible computing because they dissipate 30-60\% of all energy in (reversible) wires and repeaters \cite{Shelar13}. At room temperature, Landauer's limit amounts to 2.85$\times 10^{-21}$ J --- a very small fraction of the total, given that modern ICs dissipate 0.1-100 Watts and contain $<10^9$ logic gates. With the increasing dominance of interconnect (Section \ref{sec:space}), more energy is spent on communication than on computation. Logically-reversible computing is important for reasons other than energy --- in cryptography, quantum information processing, etc \cite{Saeedi13}.

\hide{
 An unrelated, but significant disruption occurred in the semiconductor industry due to runaway dynamic power in gates and interconnects.
 }
\subsection{Power constraints and CPUs}
\ \ \ \ {\bf The end of CPU frequency scaling.}
In 2004, Intel Corp. abruptly cancelled a 4GHz CPU project because high power density required awkward cooling technologies. Other CPU manufacturers kept clock-frequencies in the 1-6GHz range, but also resorted to multicore CPUs \cite{Borkar07}. Since dynamic circuit power grows with clock frequency and supply voltage squared \cite{Rabaey04}, energy can be saved by distributing work among slower, lower-voltage parallel CPU cores {\em if parallelization overhead is small}.
\hide{Exercise for the reader: compare this energy-time scaling to that implied by Heisenberg uncertainty \cite{Aharonov61}.}

{\bf Dark, darker, dim, gray silicon.}
A companion trend to Moore's law --- the Dennard scaling theory \cite{Bohr07} --- shows how to keep power consumption of semiconductor ICs constant while increasing their density. But Dennard scaling broke down ten years ago \cite{Bohr07}. Extrapolation of semiconductor scaling trends for CMOS --- the dominant semiconductor technology for 20 years ---  shows that the power consumption of transistors available in modern ICs reduces more slowly than their size (which is subject to Moore's law) \cite{Taylor12,Esmaielzadeh13}. To ensure the performance envelope of transistors, chip power density must be limited, and a fraction of transistors must be kept dark at any given time. Modern CPUs have not been able to use all their circuits at once, but this asymptotic effect --- termed the {\em utilization wall} \cite{Taylor12} --- will soon black out 99\% of the chip, hence the term {\em dark silicon} and a reasoned reference to the apocalypse \cite{Taylor12}. Saving power by slowing CPU cores down is termed {\em dim silicon}. Detailed studies of dark silicon \cite{Esmaielzadeh13} show similar results. To this end, executives from Microsoft and IBM have recently proclaimed an end to the era of multicore microprocessors \cite{IBM13}.
Two related trends appeared earlier: $(i)$ increasingly large IC regions remain transistor-free to aid routing and physical synthesis, to accommodate power-ground networks, \hide{and circuit modules with awkward geometries,} etc \cite{Caldwell03,Adya06} --- we call them {\em darker silicon}, $(ii)$ increasingly many gates do not perform useful computation but reinforce long, weak interconnects \cite{Saxena04} or slow down wires that are too short ---  call them {\em gray silicon}. \hide{With buffers inserted to fix short-path violations,} Today, 50-80\% of all gates in high-performance ICs are repeaters.
 \hide{Unused regions are rare in high-volume low-margin IC products, but appear increasingly frequent in large ICs that can be profitable without using every $\mu m^2$ of silicon.} 

{\bf Limits for power supply and cooling.}
 Data centers in the US consumed 2.2\% of total U.S. electricity in 2011.
 As powerplants take time to build, we cannot sustain past trends of doubled power consumption per year. It is possible to improve the efficiency of transmission lines (using high-temperature superconductors \cite{Oestergaard01}) and power conversion in datacenters, but the efficiency of on-chip power-networks may soon reach 80-90\%. Modern IC power management includes {\em clock} and {\em power gating} \cite{Borkar07}, per-core voltage scaling \cite{Pinckney13}, {\em charge recovery} \cite{Kim05} and, in recent processors, a CPU core dedicated to power scheduling. IC power consumption depends quadratically on supply voltage, which has decreased steadily for many years, but recently stabilized at 0.5-2V \cite{Rabaey04}. Supply voltage typically exceeds the {\em threshold voltage} of field-effect transistors by a safety margin that ensures circuit reliability, fast operation and low leakage. Threshold voltage depends on the thickness of gate dielectric, which reached a practical limit of several atoms (Section \ref{sec:eng}).
 Supply voltage is limited by around 200mV \cite{Meindl95} --- five times below current practice --- and simple circuits reach this limit. With slower operation, {\em near-} and {\em sub-threshold circuits} may consume 100 times less energy \cite{Dreslinski10}.
\hide{Server processors can additionally employ {\em active cooling techniques}.
  and improve efficiency through {\em cryogenic operation}.}
 Cooling technologies can improve too, but fundamental quantum limits
 bound the efficiency of heat removal \cite{Pendry83,Blencowe00,Whitney14}.

\subsection{Broader limits}
\hide{Given that alternative technologies circumvent engineering limits in Section \ref{sec:eng}, we look for limits to computation beyond CMOS.}
The study in \cite{ZhirnovCHB03} explores a general {\em binary-logic switch model} with binary states represented by two {\em quantum wells separated by a potential barrier}. Representing information by electric charge requires energy for binary switching and thus limits the logic-switching density, if a significant fraction of the chip can switch simultaneously. To circumvent this limit, one can encode information in {\em spin-states, photon polarizations, super-conducting currents}, or {\em magnetic flux}, noting that these carriers have already been in commercial use. Spin-states are particularly attractive because they promise high-density nonvolatile storage \cite{Wolf01} and scalable interconnects \cite{Naeemi14}.
More powerful limits are based on the amount of material in Earth's crust (where silicon is the second most common element after oxygen), on atomic spacing (Section \ref{sec:eng}), radii and energies, bandgaps, as well as the wavelength of the electron. We are currently using only a tiny fraction of Earth's mass for computing, and yet various limits
could be circumvented if new particles are discovered.
Beyond atomic physics, some limits rely on basic constants:  the speed of light, the gravitational constant, the quantum (Planck) scale, the Boltzmann constant, etc.
Lloyd \cite{Lloyd00}, as well as Kraus \cite{Krauss04} extend well-known bounds by Bremermann and Bekenstein, and give Moore's law 150 and 600 years, respectively. These results are too loose to obstruct the performance of practical computers. In contrast, current consensus estimates from the ITRS \cite{ITRS} give Moore's law only 10-20 years, due to both {\em technological} and {\em economic} considerations \cite{Chien13}.

\section{Asymptotic space-time limits}
\label{sec:space}
   Engineering limits for deployed technologies can often be circumvented, while first-principles limits on energy and power are very loose. Reasonably tight limits are rare.
{\bf Limits to parallelism.}
Suppose we wish to compare a parallel and sequential computer built from the same units, to argue that a new parallel algorithm is many times faster than the best sequential algorithm (the same reasoning applies to logic gates on an IC). Given $N$ parallel units and an algorithm that runs $K$ times faster on sufficiently large inputs, one can {\em simulate} the parallel system on the sequential system by dividing its time between $N$ computational slices. Since this simulation is roughly $N$ times slower, it runs $K/N$ times faster than the original sequential algorithm. If this algorithm was best possible, we have $K\leq N$. The bound is reasonably tight in practice for small $N$ and can be violated slightly since $N$ CPUs include more CPU cache, but such violations do not justify parallel algorithms --- one could instead buy/build one CPU with a larger cache.
Such linear speedup is optimistically assumed for {\em the parallelizable component} in the 1988 Gustafson's law that suggests scaling the number of processors with input size (as illustrated by instantaneous Web search queries over massive data sets) \cite{Padua11}.
Also in 1988, Fisher \cite{Fisher88} employed {\em asymptotic runtime estimates} instead of numerical limits and avoided the breakdown into parallel and sequential runtime components, assumed in Amdahl's \cite{Amdahl13} and Gustafson's laws \cite{Padua11}. Asymptotic estimates neglect leading constants and offer a powerful way to capture nonlinear phenomena occurring at large scale.
\hide{(small-scale computations are irrelevant to scaling limits).}

Fisher \cite{Fisher88} assumes a sequential computation with $T(n)$ elementary steps for input of size $n$, and limits the performance of its parallel variants that can use an unbounded $d$-dimensional grid of finite-size computing units (electrical switches on a semiconductor chip, logic gates, CPU cores, etc) communicating at a finite speed, say, bounded by the speed of light. We highlight only one aspect of this four-page work --- the parallel computation requires $\Omega(\sqrt[d+1]{T(n)})$ steps. This result undermines the $N$-fold speedup assumed in Gustafson's law for $N$ processors on appropriately sized input data \cite{Padua11}. A more realistic speedup from $\sim n^k$ to $\sim \log n$ can be achieved in an abstract model of computation for matrix multiplication and fast Fourier transforms. But not in physical space \cite{Fisher88}. Surprising as it may seem, after reviewing many loose limits to computation, we have identified a reasonably tight limit (the impact of I/O --- a major bottleneck today --- is also covered in\cite{Fisher88}). Indeed, many parallel computations today (excluding multimedia processing and Web search) are limited by several forms of communication and synchronization, including network and storage access. The billions of logic gates and memory elements in modern ICs are linked by up to 16 levels of wires (Figure \ref{fig:wires}), longer wires are segmented by repeaters. Most of the physical volume and circuit delay are attributed to interconnect \cite{Shelar13}. This is relatively new, as gate delays were dominant until 2000 \cite{ITRS}, but wires get slower relative to gates at each new technology node. This uneven scaling has compounded in ways that would surprise Turing and von Neumann --- a single clock cycle is now far too short for a signal to cross the entire chip, and even the distance covered in 200 ps (5 GHz) at light-speed is close to chip size. Yet, most electrical engineers and computer scientists continue to focus on gates.

{\bf Implications to 3D ICs and other emerging technologies.}
 The promise of 3D integration for improving IC performance can be contrasted with  technical obstructions to its industry adoption. To derive limits on possible improvement, we use the result from \cite{Fisher88} sensitive to the dimension of the physical space: a sequential computation with $T(n)$ steps requires $\Omega(\sqrt[3]{T(n)})$ steps in 2D and $\Omega(\sqrt[4]{T(n)})$ in 3D. Letting $t=\sqrt[3]{T(n)}$, shows that 3D integration asymptotically reduces $t$ to $t^{3/4}$ --- a significant but not dramatic speedup. This speedup requires an unbounded number of 2D device layers, otherwise there is no asymptotic speedup \cite{Mak12}. For 3D ICs with 2-3 layers, the main benefits of 3D IC integration today are in improving manufacturing yield, improving I/O bandwidth, and combining 2D ICs that are optimized for random logic, dense memory, FPGA, analog, MEMS, etc. Ultra-high density CMOS logic ICs with {\em monolithic} 3D integration \cite{Lee12} suffer higher routing congestion than traditional 2D ICs.
 Emerging technologies promise to improve device parameters, but often remain limited by scale, faults, and interconnect, e.g., {\em quantum dots} enable Terahertz switching but hamper nonlocal communication \cite{Sherwin99}. CNT-FETs \cite{Shulaker13} leverage extraordinary carrier mobility in semiconducting carbon nanotubes to use interconnect more efficiently by improving drive strength, while reducing supply voltage. Emerging interconnects include {\em silicon photonics}, shown by Intel in 2013 \cite{Simonite13} as a 100Gb/s replacement of copper cables connecting adjacent chips. It promises to reduce power consumption and form factor. Quantum physics alters the nature of communication with Einstein's ``spooky action at a distance'' facilitated by entanglement  \cite{Nielsen11}.
 However, the flows of information and entropy are subject to quantum limits \cite{Pendry83,Blencowe00}. Several quantum algorithms run asymptotically faster than best conventional algorithms \cite{Nielsen11}, but fault-tolerance overhead offsets their potential benefits in practice, and empirical evidence of quantum speedups has not been compelling so far \cite{Ronnow14,Shin14}. Several stages in the development of quantum information processing remain challenging \cite{Devoret13}, and the surprising difficulty of {\em scaling up} reliable quantum computation could stem from limits on {\em communication} and {\em entropy} \cite{Pendry83,Blencowe00,Nielsen11}. In contrast, Lloyd \cite{Lloyd00} notes that {\em individual} quantum devices now approach energy limits for switching, whereas nonquantum devices remain orders of magnitude away. This suggests an obstacle to simulating quantum physics on conventional computers (abstract models aside). 
 In terms of computational complexity though, quantum computers {\em cannot} attain significant advantage for many problem types \cite{Aaronson04,Jain10,Nielsen11}. Such lack of {\em consistent general-purpose speedup} limits the benefits of several emerging technologies in mature applications with diverse algorithmic steps, e.g., computer-aided design and Web search. Accelerating one step usually does not greatly speed up the entire application, as noted by Amdahl in 1967 \cite{Amdahl13}. Figuratively speaking, {\em the most successful computers are designed for the decathlon, rather than for sprint only}.%

\section{Complexity-theoretic limits}
\label{sec:nonphys}
Section \ref{sec:space} enabled tighter limits by neglecting energy and using asymptotic rather than numeric bounds --- a more abstract model focuses on the impact of scale, and recurring trends quickly overtake one-off device-specific effects.
\hide{In practice, when a large programmable chip or a supercomputer become available, end-users try to neglect energy considerations for as long as possible, but cannot ignore runtime and must often account for interconnect delays.
}
Next, we neglect spatial effects and focus on the nature of computation in an abstract model (used by software engineers) that represents computation by elementary steps with input-independent runtimes.
Such limits survive many improvements in computer technologies, and are often stronger for specific problems. For example, the best-known algorithms for multiplying large numbers are only slightly slower than reading the input (an obvious speed limit), but only in the asymptotic sense --- for numbers with $<$1000 bits, those algorithms lag behind simpler algorithms in actual performance.
\hide{If the reader questioned our logic in Section \ref{sec:space}, our further reasoning may also seem odd.}
To focus on what matters, we now do not just track asymptotic worst-case complexity of best algorithms for a given problem, but merely distinguish {\em polynomial} asymptotic growth from {\em exponential}. Limits formulated in such crude terms (unsolvability in polynomial time {\em on any computer}) are powerful \cite{Sipser12}: the hardness of number-factoring underpins Internet commerce, while the P$\neq$NP conjecture explains the lack of satisfactory, scalable solutions to important algorithmic problems, e.g., in optimization and verification of IC designs \cite{Fortnow09}. A similar conjecture P$\neq$NC seeks to explain why many algorithmic problems that can be {\em solved} efficiently have not {\em parallelized} efficiently \cite{Markov13}. Most of these limits have not been proven. Some can be circumvented by using radically different physics, e.g., quantum computers solve number factoring in polynomial time (in theory). But quantum computation does not affect P$\neq$NP \cite{Aaronson05}. The lack of proofs, despite heavy empirical evidence, requires faith and is an important limitation of many nonphysical limits to computing. This faith is not universally shared --- Donald Knuth argues\footnote{See Question 17 in \url{http://www.informit.com/articles/article.aspx?p=2213858}} that P=NP would not contradict anything we know today. A rare {\em proven} result by Turing (also invulnerable to quantum physics) states that checking if a given program ever halts is {\em undecidable}: no algorithm solves this problem in all cases regardless of runtime. Yet, software developers solve this problem during peer code reviews, and computer science teachers --- when grading exams in programming courses. {\em Worst-case analysis} is another limitation of nonphysical limits to computing, but suggests potential gains through approximation and specialization. For some NP-hard optimization problems, such as the {\em Euclidean Travelling Salesman Problem} (EucTSP), polynomial-time approximations exist, but in other cases, such as the {\em maximum clique problem}, accurate approximation is as hard as finding optimal solutions \cite{Vazirani02}. For some important problems and algorithms, such as the Simplex algorithm for {\em linear programming}, few inputs lead to exponential runtime, and minute perturbations reduce runtime to polynomial \cite{Spielman01}.

\section{Conclusions}
\label{sec:conc}
 The death march of Moore's law \cite{Cavin12,Chien13} invites discussions of fundamental limits and alternatives to silicon semiconductors \cite{Shulaker13}. Near-term constraints invariably tie to {\em costs} and {\em capital}, but are explained away by new markets for electronics, increasing Earth population, and growing world economy \cite{Chien13}. Such economic pressures emphasize the value of {\em computational universality} and broad applicability of IC architectures to solve multiple tasks under conventional environmental conditions. In a likely scenario, only CPUs, GPUs, FPGAs and dense memory ICs will remain viable at the end of Moore's law, while specialized circuits will be manufactured with less advanced technologies.
 Indeed, memory chips have lead Moore scaling by leveraging their simpler structure, modest interconnect, and more controllable manufacturing, but their scaling is slowing down \cite{Chien13}. The decelerated scaling of CMOS ICs still outperforms the scaling of the most viable emerging technologies.
 Empirical scaling laws describing the evolution of computing are well-known \cite{Getov13}. In addition to Moore's law, Dennard scaling, as well as Amdahl's and Gustafson's laws reviewed earlier, Metcalfe's law \cite{Metcalfe13} states that the value of a computer network, such as the Internet or Facebook, scales as the number of user-to-user connections that can be formed. Grosch's law \cite{Ryan13} ties $N$-fold improvements in computer performance to $N^2$-fold cost increases (in equivalent units). Applying it in reverse, we can estimate acceptable performance of cheaper computers. But such laws only capture {\em ongoing scaling} and will break down in the future.
 \hide{Hence, the challenge to understand obstacles to scaling and limits to computation.}

 The {\em roadmapping process} represented by the International Technology Roadmap for Semiconductors (ITRS) \cite{ITRS} relies on consensus estimates and works around engineering obstacles. It tracks improvements in materials and tools, collects best practices and outlines promising design strategies. As suggested in \cite{Meindl95,Davis01}, it can be enriched by analysis of limits. We additionally focus on how closely such limits can be approached. Aside from historical ``wrong turns'' recalled in Sections \ref{sec:eng} and \ref{sec:energy}, we find interesting effects when examining the tightness of individual limits. While energy-time limits are most critical in computer design \cite{ITRS,Wenisch12}, space-time limits appear tighter \cite{Fisher88} and capture bottlenecks formed by interconnect and communication. They suggest optimizing gate locations and sizes, and placing gates in three dimensions. One can also adapt algorithms to spatial embeddings \cite{Bachrach07,Rosenbaum12} and seek space-time limits. But the gap between current technologies and energy-time limits hints at greater rewards. {\em Charge recovery} \cite{Kim05}, {\em power management} \cite{Borkar07}, voltage scaling \cite{Pinckney13},
 and {\em near-threshold computing} \cite{Dreslinski10} reduce energy waste. Optimizing algorithms and circuits simultaneously for energy and spatial embedding \cite{Patil07} gives biological systems an edge (from the 1D worm {\em C. elegans} with 302 neurons to the 3D human brain with 86 billion neurons) \cite{Cavin12}. Yet, using mass-energy to compute can be a veritable {\em nuclear option}. In a 1959 talk, which predated Moore's law, Richard Feynman suggested that there was ``plenty of room at the bottom,'' forecasting the miniaturization of electronics. Today, with relatively little physical room left, {\em there is plenty of {\em energy} at the bottom}. If this energy is tapped for computing, how can resulting heat be removed? Recycling heat into mass or electricity seems ruled out by limits to energy conversion and the acceptable thermal envelope.

 Technology-specific limits for modern computers tend to express tradeoffs, especially for systems with conflicting performance parameters and properties \cite{CAP12}. Little is known about limits on {\em design technologies}. Given that large-scale complex systems are often designed and implemented hierarchically \cite{Caldwell03} with multiple levels of abstraction, it would be valuable to capture losses incurred at abstraction boundaries and between levels of design hierarchies. It is common to estimate resources required for a subsystem and then implement the subsystem to satisfy resource budgets. Underestimation is avoided because it leads to failures, but overestimation results in overdesign. Inaccuracies in estimation and physical modeling also lead to losses during optimization, especially in the presence of uncertainty. Clarifying engineering limits gives hope to circumvent them.

 Technology-agnostic limits look simple and have had significant impact in practice,
 for example Aaronson explains why NP-hardness is unlikely to be circumvented by through physics \cite{Aaronson05}.
  Limits to {\em parallel computation} became prominent after CPU speed levelled off ten years ago. They suggest using faster interconnect \cite{Davis01}, local computation that reduces communication \cite{Demmel13}, time-division multiplexing of logic \cite{Hatfill10}, architectural and algorithmic techniques \cite{Dror11}, solving larger problem instances, and altering applications to embrace parallelism \cite{Padua11}. John Gustafson advocates a {\em natural selection}: the survival of applications fittest for parallelism. In another twist, the performance and power consumption of industry-scale distributed systems is often described by probability distributions, rather than single numbers \cite{Dean13,Barroso13}, making it harder to even formulate appropriate limits.
  We also cannot yet formulate fundamental limits related to the complexity of the software-development effort, the efficiency of CPU caches \cite{Jouppi11}, and computational requirements of incremental functional verification, but we have noticed that many known limits are either loose or can be circumvented, leading to {\em secondary limits}. To wit, the $P\neq NP$ limit is worded in terms of worst-case rather than average-case performance, and has not been proven despite heavy evidence. Researchers have ruled out entire categories of proof techniques as insufficient to complete such a proof \cite{Fortnow09,Aaronson09}. While esoteric, such {\em tertiary limits} can be effective in practice --- in August 2010, they helped researchers quickly invalidate Vinay Deolalikar's highly-technical attempt at proving $P\neq NP$. On the other hand, the correctness of lengthy proofs for some key results could not be established with acceptable level of certainty by reviewers, prompting efforts in verifying mathematics by computation \cite{Avigad14}.

  In summary, we have reviewed what is known about limits to computation, including
  existential challenges arising in the sciences, design and optimization challenges arising in engineering, as well as current state of the art. These categories are closely linked due to the rapid pace of technology development. When a specific limit is approached and obstructs progress, understanding its assumptions is a key to circumventing it. Some limits are hopelessly loose and can be ignored,
  while other limits remain conjectured based on empirical evidence and may be very
  difficult to establish rigorously. Such {\em limits on limits to computation} deserve further study.

 \noindent
 {\bf Acknowledgments.} This work was supported in part by the Semiconductor Research Corporation (SRC) Task 2264.001 (funded by Intel and IBM), US Airforce Research Laboratory Award FA8750-11-2-0043, and US National Science Foundation (NSF) Award 1162087.

\end{document}